\documentclass[11pt,twoside]{article} 
\usepackage{cspm-asp2006}
\usepackage{epsfig,graphicx,natbib,url}  
\usepackage{lscape} 
\begin{document}
\setcounter{page}{441}

\markboth{Ka\v{s}parov\'{a} et al.}{H$\alpha$ with Particle Beams} 
\title{H$\alpha$ with Heating by Particle Beams} 
\author{J. Ka\v{s}parov\'{a}$^{1}$,
	M. Varady$^{1,2}$,
	M. Karlick\'{y}$^{1}$,
	P. Heinzel$^{1}$
	and 
	Z. Moravec$^{2}$
	} 
\affil{
	$^{1}$ Astronomical Institute AS, Ond\v{r}ejov, Czech Republic\\
	$^{2}$ Department of Physics, J. E. Purkinje University, Czech Republic
	} 

\begin{abstract} 
Using 1D NLTE radiative hydrodynamics we model the influence of the
particle beams on the H$\alpha$ line profile treating the beam
propagation and the atmosphere evolution self-consistently.  We focus
on the influence of the non-thermal collisional rates and the return
current.  Based on our results, we propose a diagnostic method for
determination of the particle beam presence in the formation regions
of the H$\alpha$ line.
\end{abstract}

\section{Introduction}
Some of the flare models assign a fundamental role to the high energy
particle beams in the flare energy transport.  As the beams interact
with the ambient plasma, their energy is dissipated and transformed
mainly into the thermal energy of the transition region and
chromosphere plasma.  Several models studied electron and proton beams
as heating agents (e.g., \cite{kasparova-ma89,kasparova-em98}) as well
as their influence on spectral line profiles
(e.g., \cite{kasparova-al05}).  The propagation of electron beams is
inevitably connected with the so-called return current (RC,
\cite{kasparova-oo90}) which also contributes to the beam energy
dissipation.  Besides the heating, the beams influence atomic level
populations of the ambient plasma via collisional excitation and
ionization.  The work presented here concentrates on these two effects
which are commonly neglected in the flare modeling and assesses their
importance on the formation of H$\alpha$ line in solar flares.

\section{Model}
The beam propagation and energy deposition is modelled by a test
particle approach consistently with the hydrodynamics of the
atmosphere and NLTE radiative transfer in the transition region,
chromosphere, and photosphere.  Details of the model and the methods
used are described in \citet{kasparova-ka05c}.

We study the response of quiet Sun atmosphere (VAL C from
\cite{kasparova-ve81}) to beam pulses of short duration, 1~s with
sinus-like time modulation, and power-law energy spectrum with
$\delta=3$.

\section{Hydrodynamics and Beam Propagation}
The model takes into account Coulomb collisions of the beam with
ambient neutrals and electrons, scattering of beam electrons, and
optionally RC for the case of an electron beam (return current is a
factor of $E_{\rm e}/E_{\rm p}$ lower for protons carrying the same
power as electrons (\cite{kasparova-br90}) and is neglected).  The
return current is included in a runaway approximation assuming
$\alpha=0.1$, i.e.\ 10\% of ambient electrons carry RC. For details
and other approximations of RC in solar atmosphere conditions see
\citet{kasparova-va05c,kasparova-va07}.

\begin{figure}
  \centering
  \includegraphics[width=0.80\textwidth]{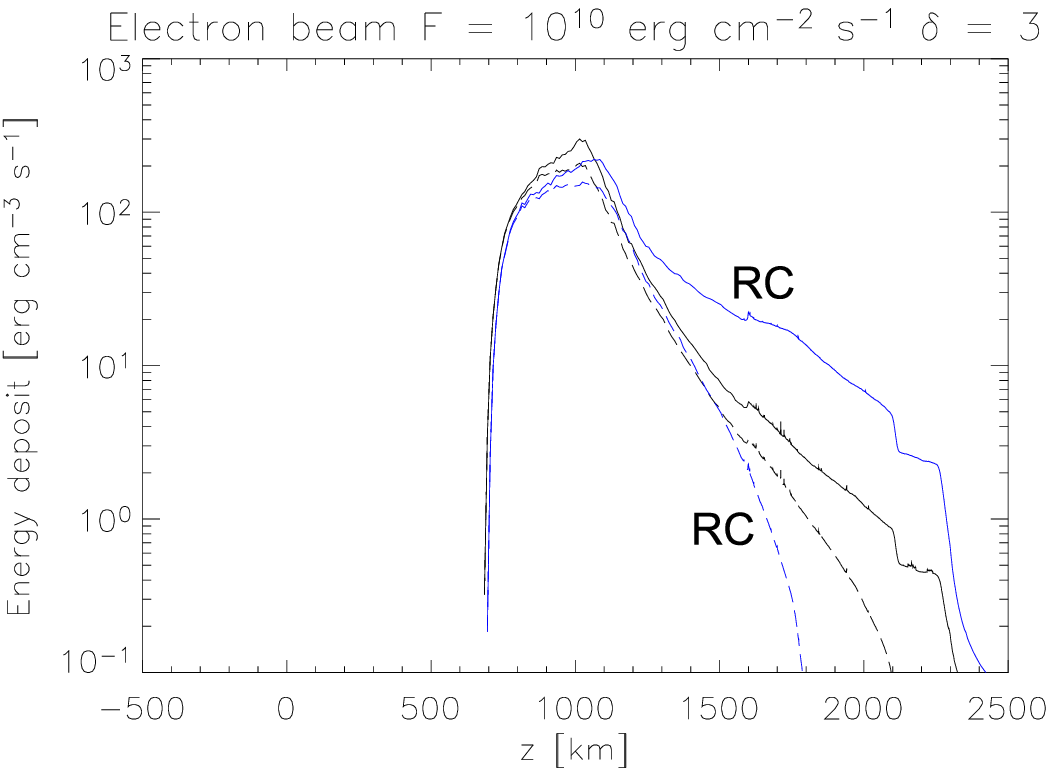}\hspace{1cm}
  \includegraphics[width=0.80\textwidth]{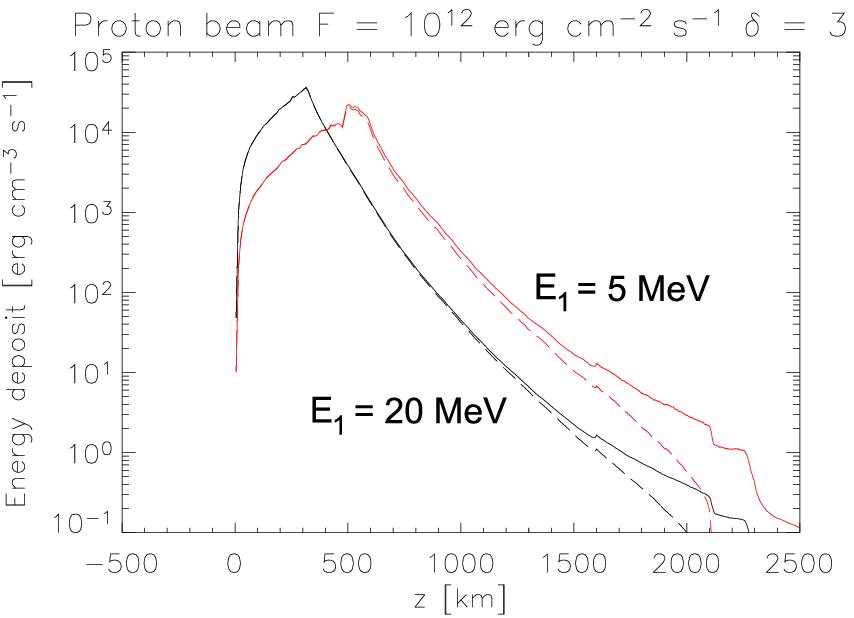}
  \caption[]{\label{kasparova-fig:deposit}
  Total energy deposit (solid) and energy deposit to hydrogen (dashed) at beam flux maximum. 
  {\em Top\/}: electron beam with and without RC included.
  {\em Bottom\/}: proton beam with $E_1=5$~MeV and $E_1=20$~MeV.
}
\end{figure}

\begin{figure}
  \centering
  \includegraphics[height=0.37\textwidth]{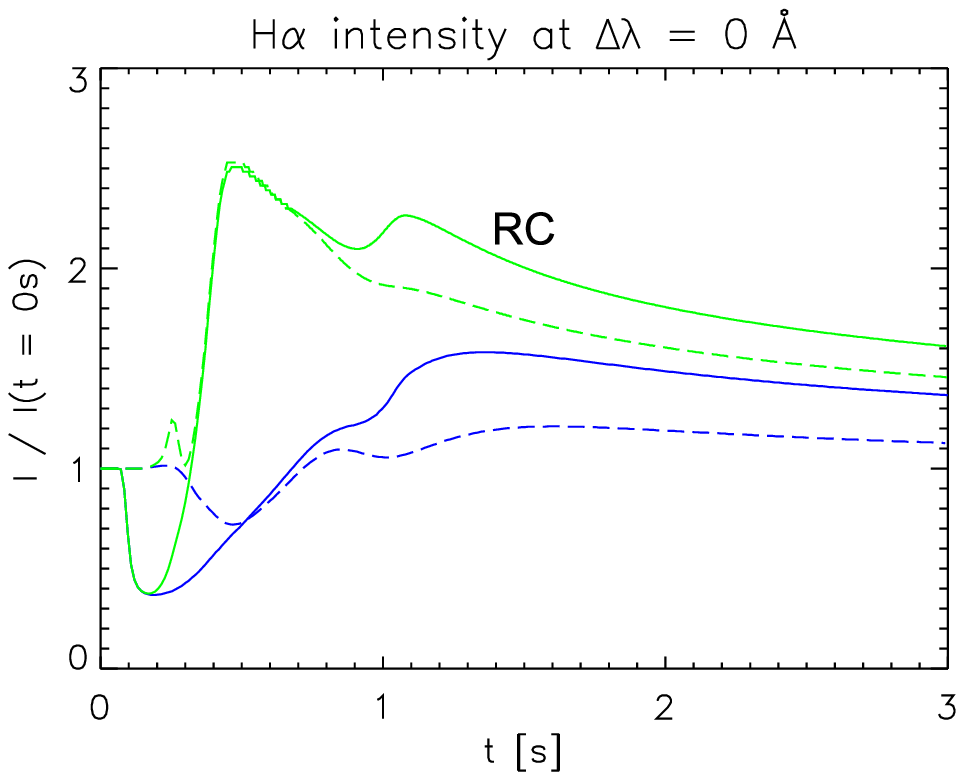}\hspace{1cm}
  \includegraphics[height=0.37\textwidth]{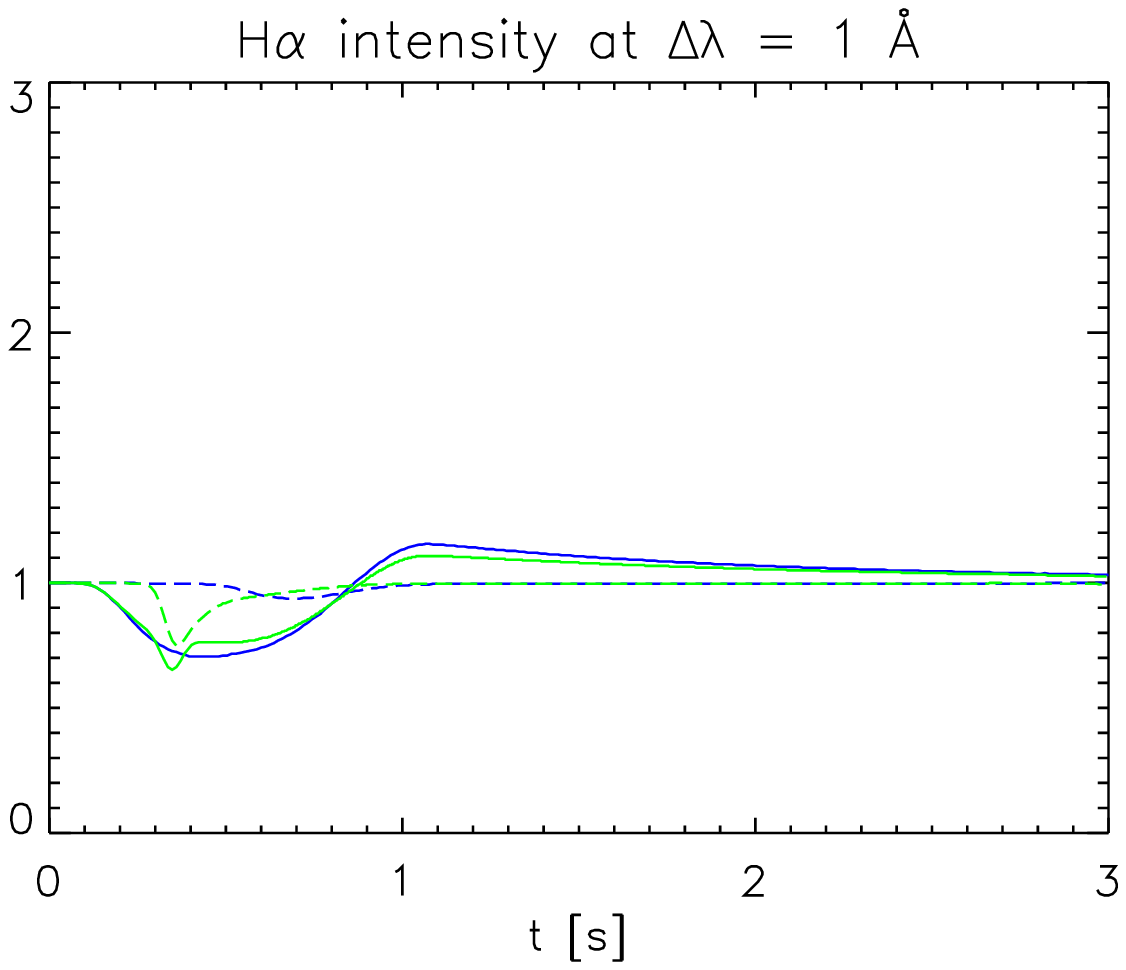}\\
  \includegraphics[height=0.37\textwidth]{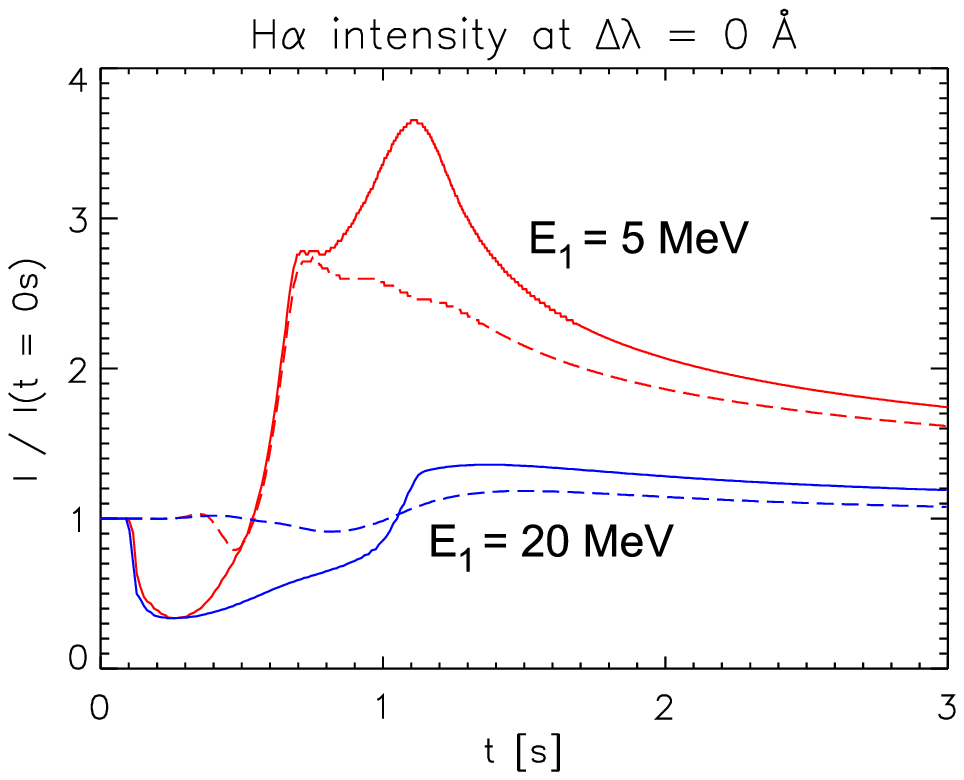}\hspace{1cm}
  \includegraphics[height=0.37\textwidth]{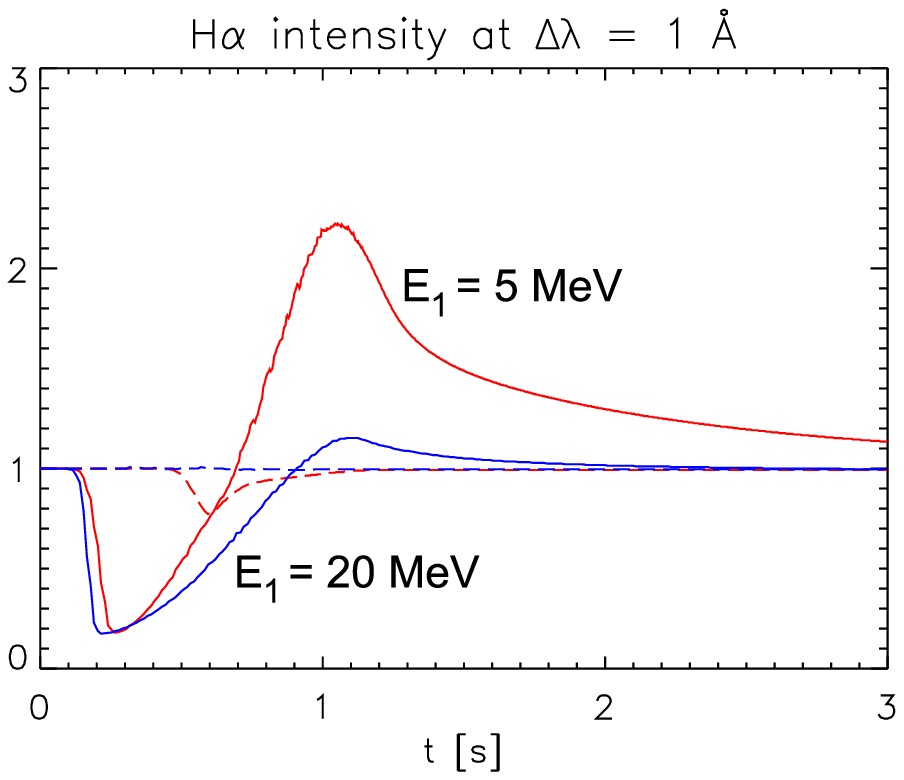}
  \caption[]{\label{kasparova-fig:prof}
  Time evolution of the H$\alpha$ line center (left column) and wing (right column) intensities.
  {\em Top\/}: electron beam for $F=10^{10}$~erg\,cm$^{-2}$\,s$^{-1}$ and $\delta=3$ with and without RC. 
  {\em Bottom\/}: proton beam for $F=10^{12}$~erg\,cm$^{-2}$\,s$^{-1}$, $\delta=3$, and $E_1=5,\ 20$~MeV. 
  The solid curves refer to cases with $C^{\rm nt}$ included, dashed lines to cases without $C^{\rm nt}$.
  }		
\end{figure}

\begin{figure}
  \centering
  \includegraphics[width=0.49\textwidth]{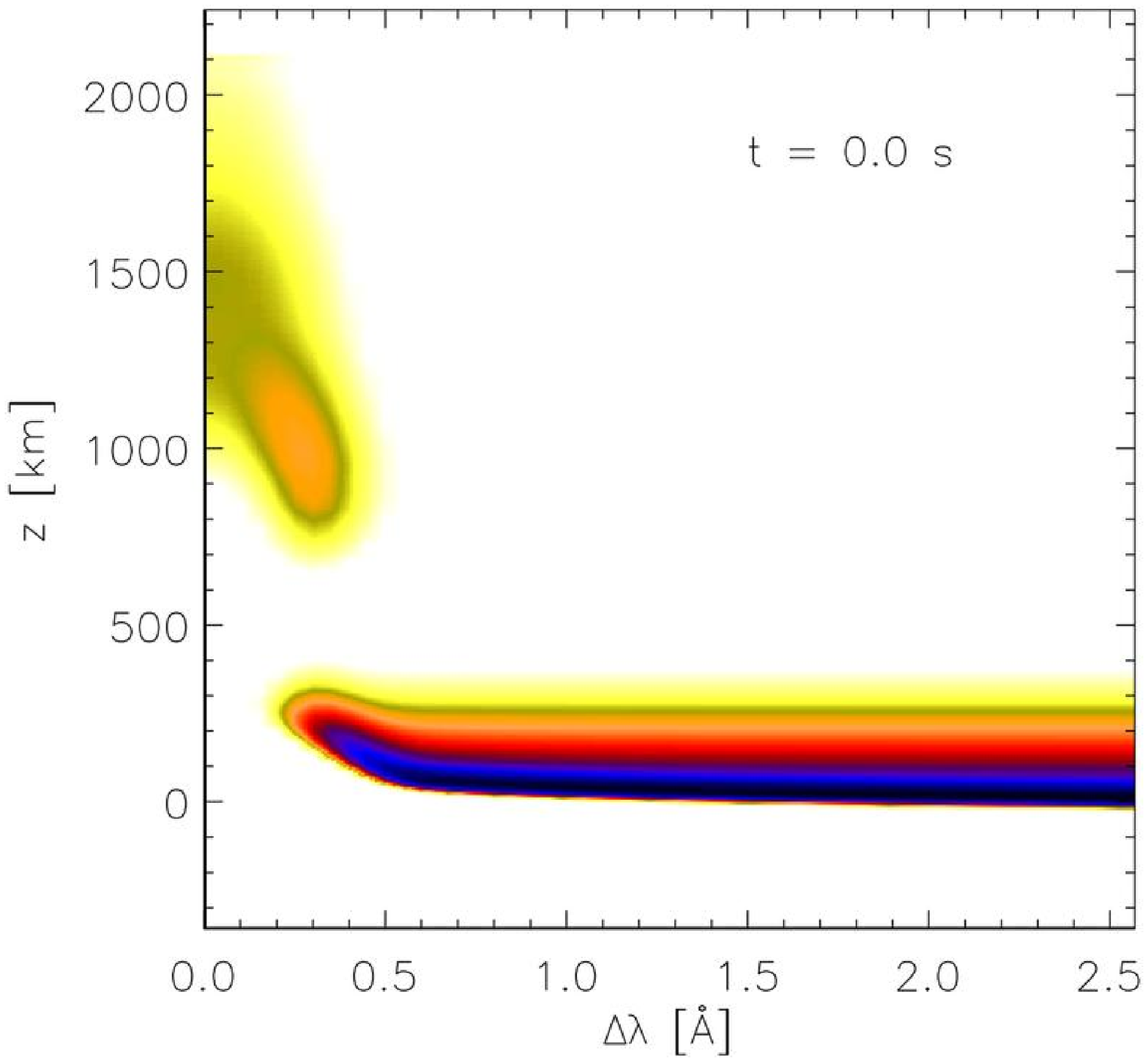}
  \includegraphics[width=0.49\textwidth]{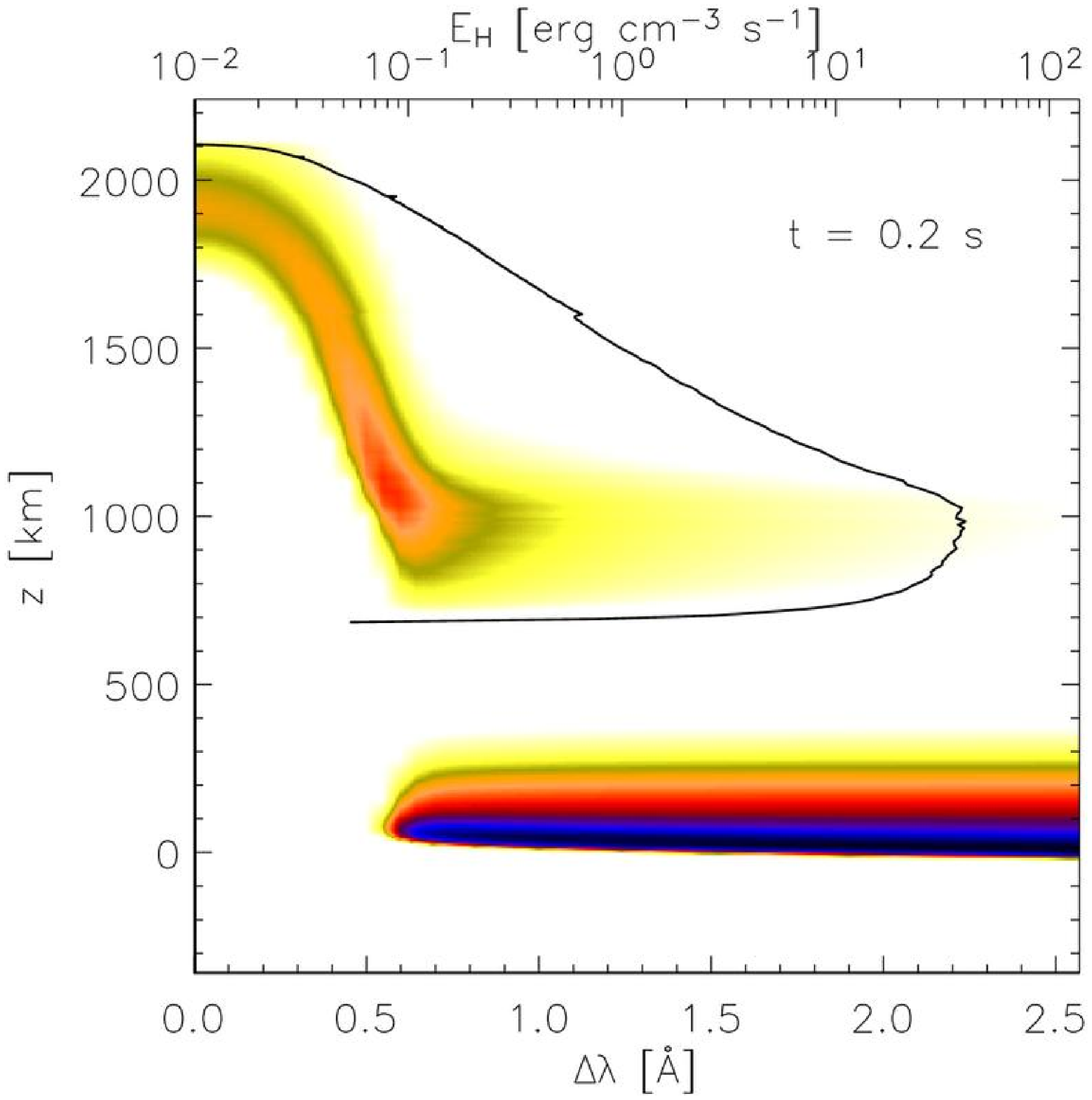}	
  \includegraphics[width=0.49\textwidth]{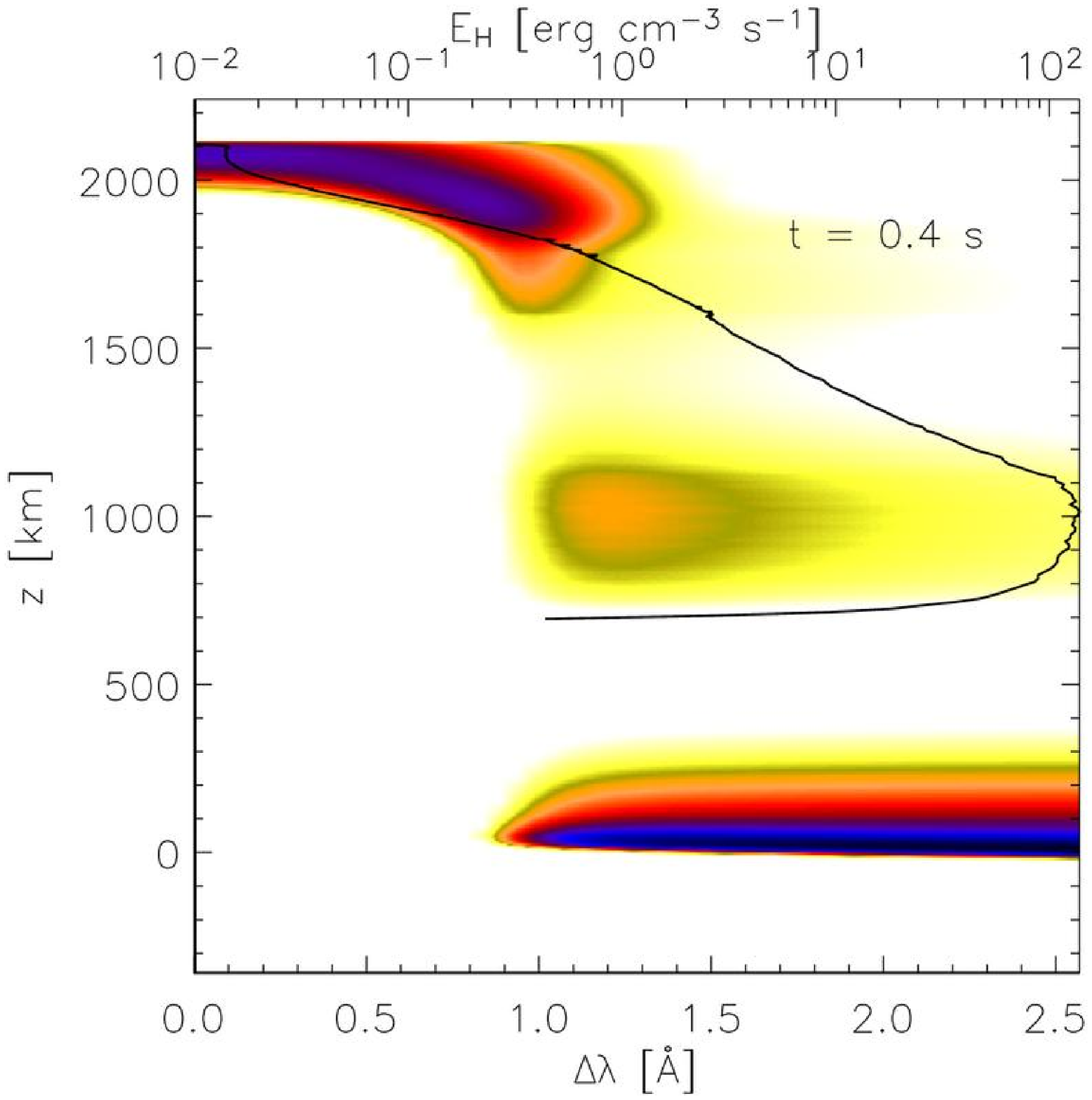}
  \includegraphics[width=0.49\textwidth]{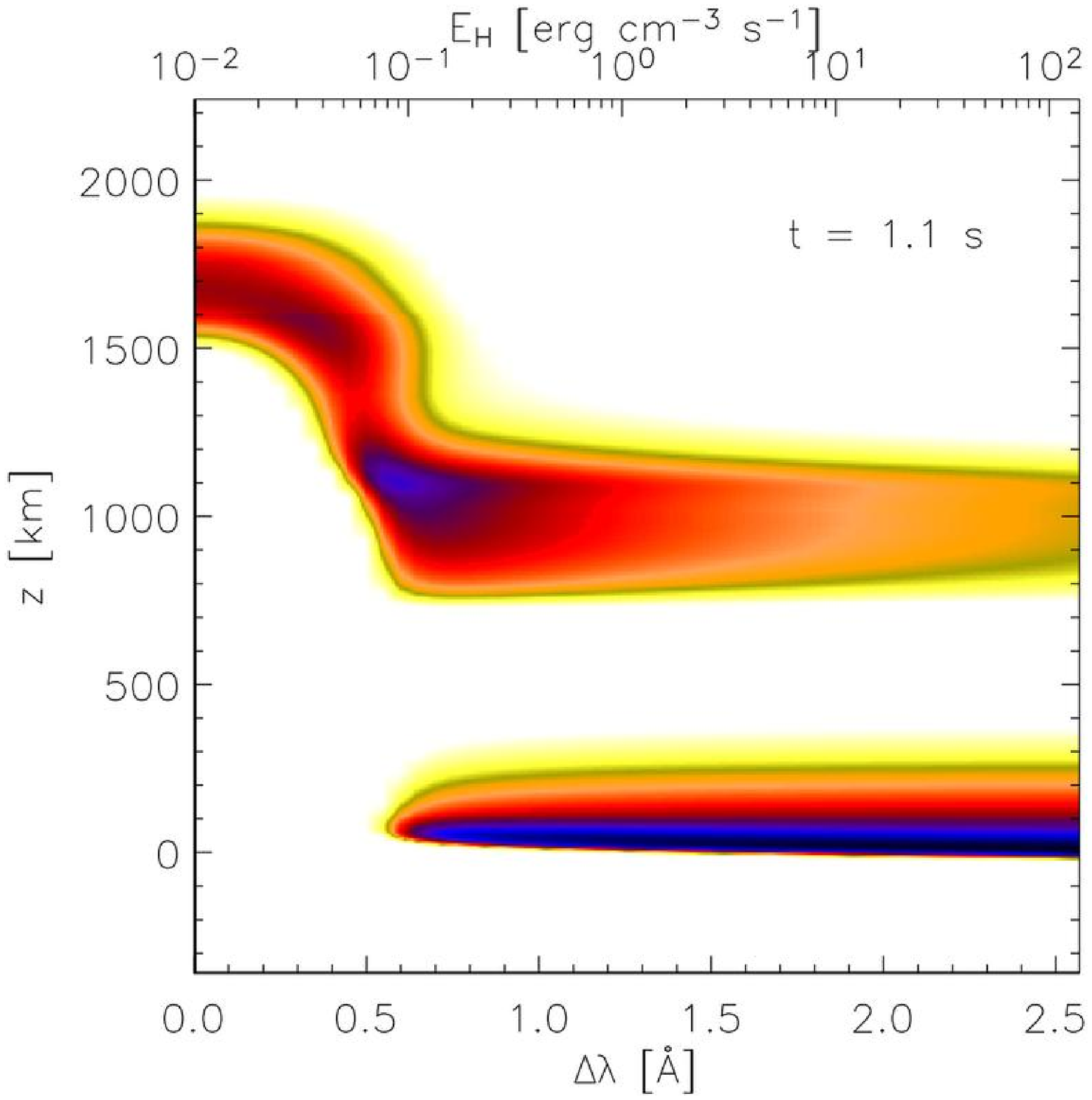}
  \caption[]{\label{kasparova-fig:cfel} 
  H$\alpha$ contribution function $CF$ for an electron beam. 
  RC and $C^{\rm nt}$ were included. 
  Solid curve: energy deposit on hydrogen.
}
\end{figure}

\begin{figure}
  \centering
  \includegraphics[width=0.46\textwidth]{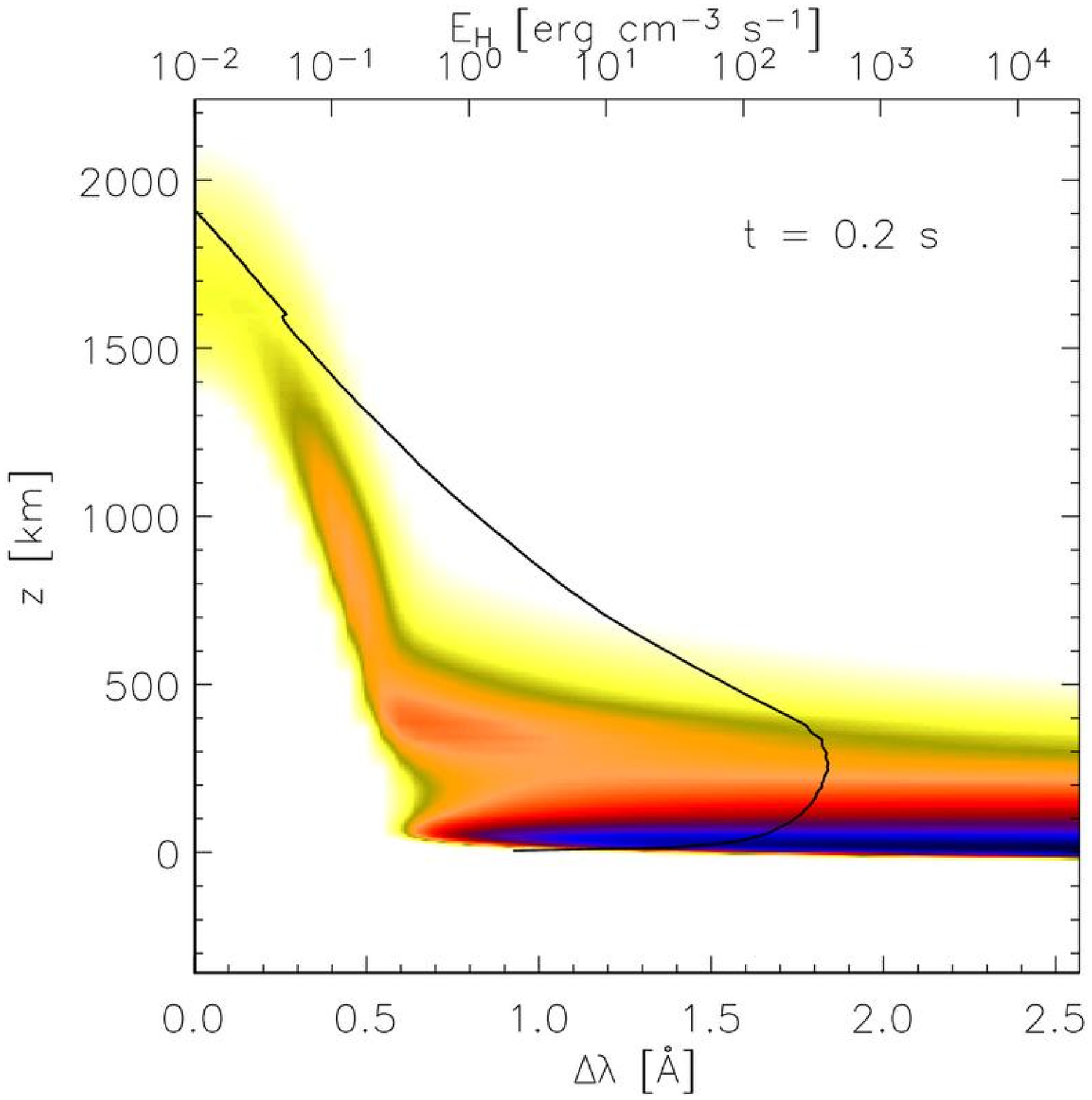}
  \includegraphics[width=0.46\textwidth]{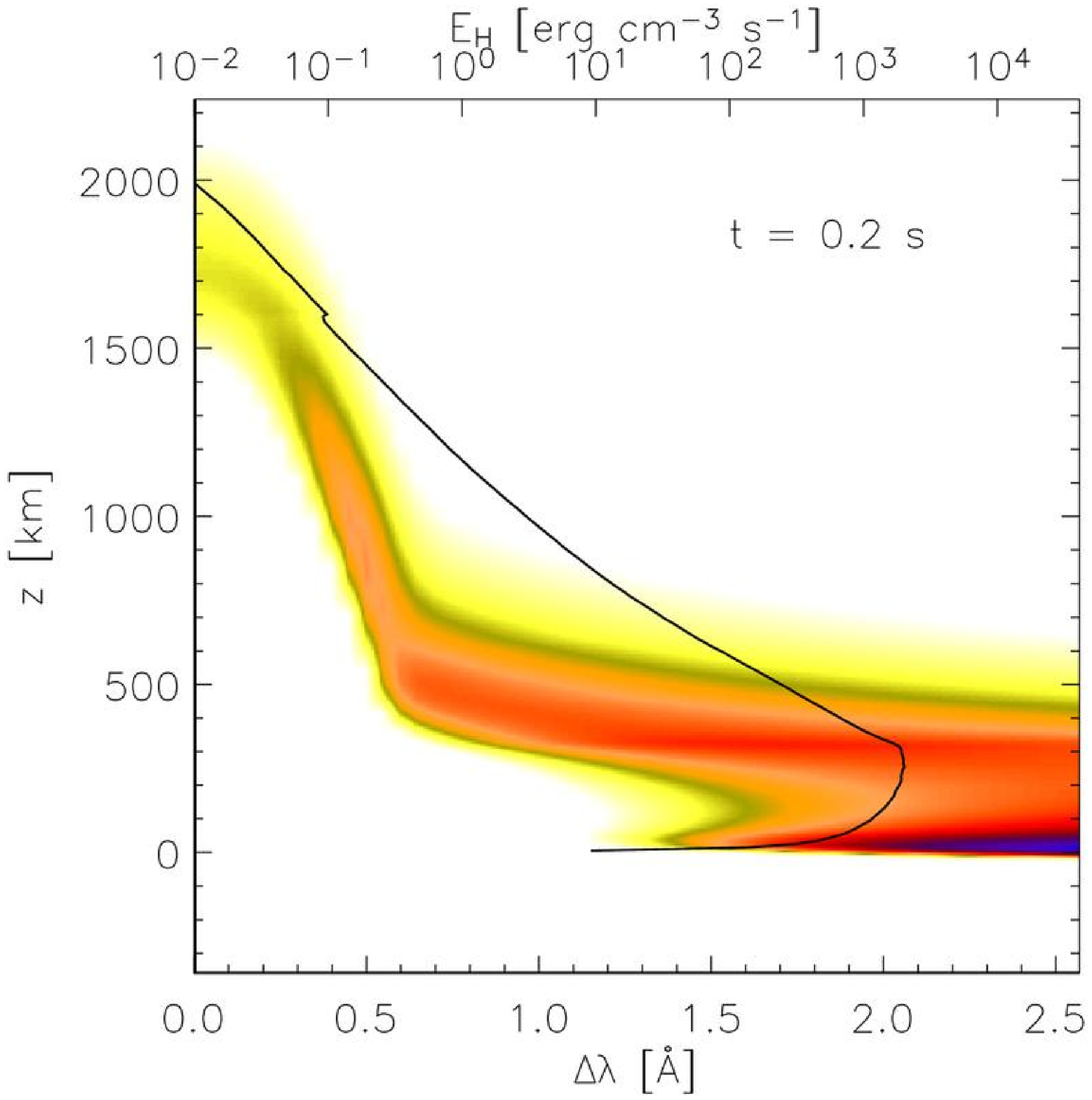}\\
  \includegraphics[width=0.46\textwidth]{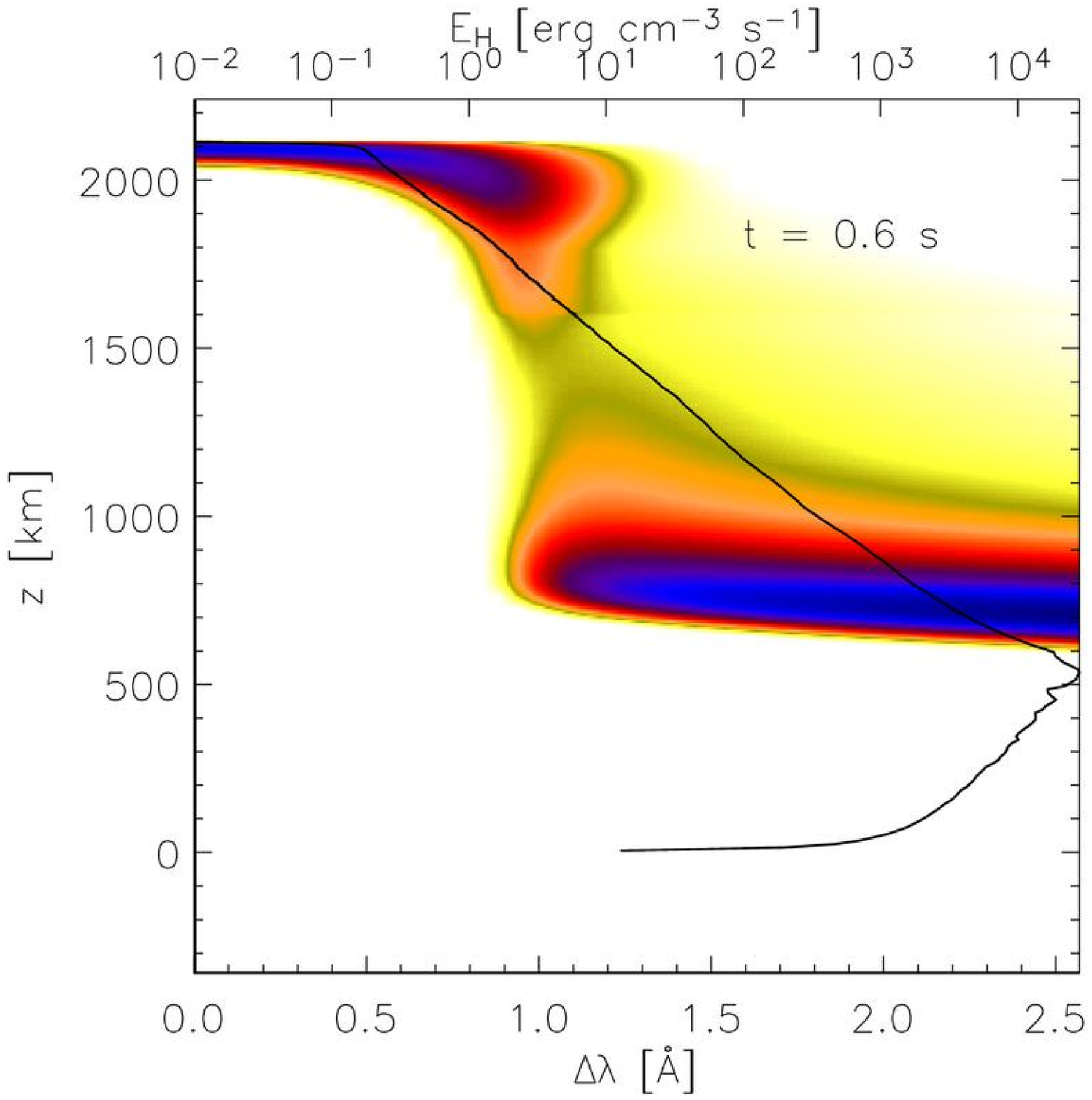}
  \includegraphics[width=0.46\textwidth]{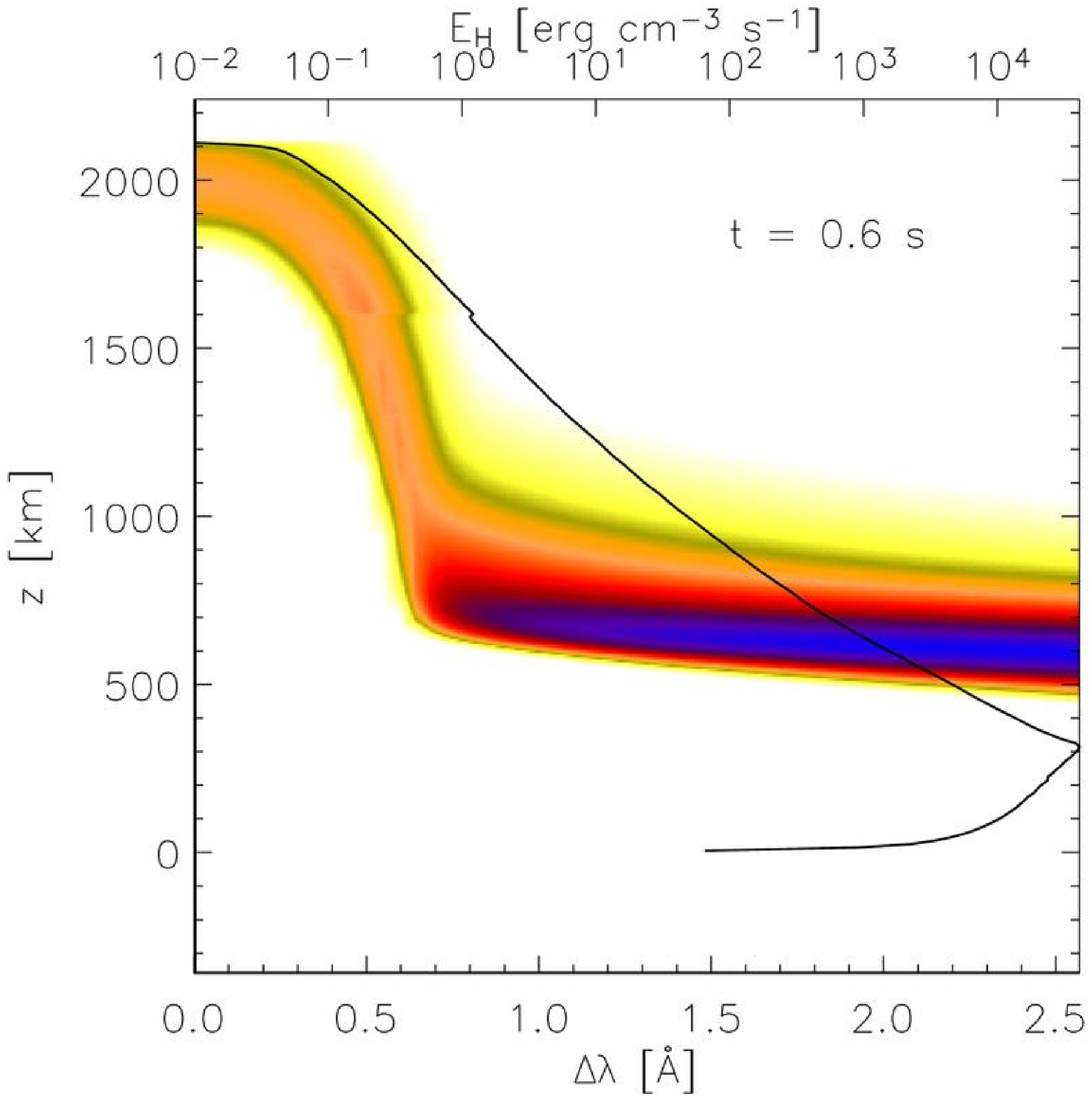}\\
  \includegraphics[width=0.46\textwidth]{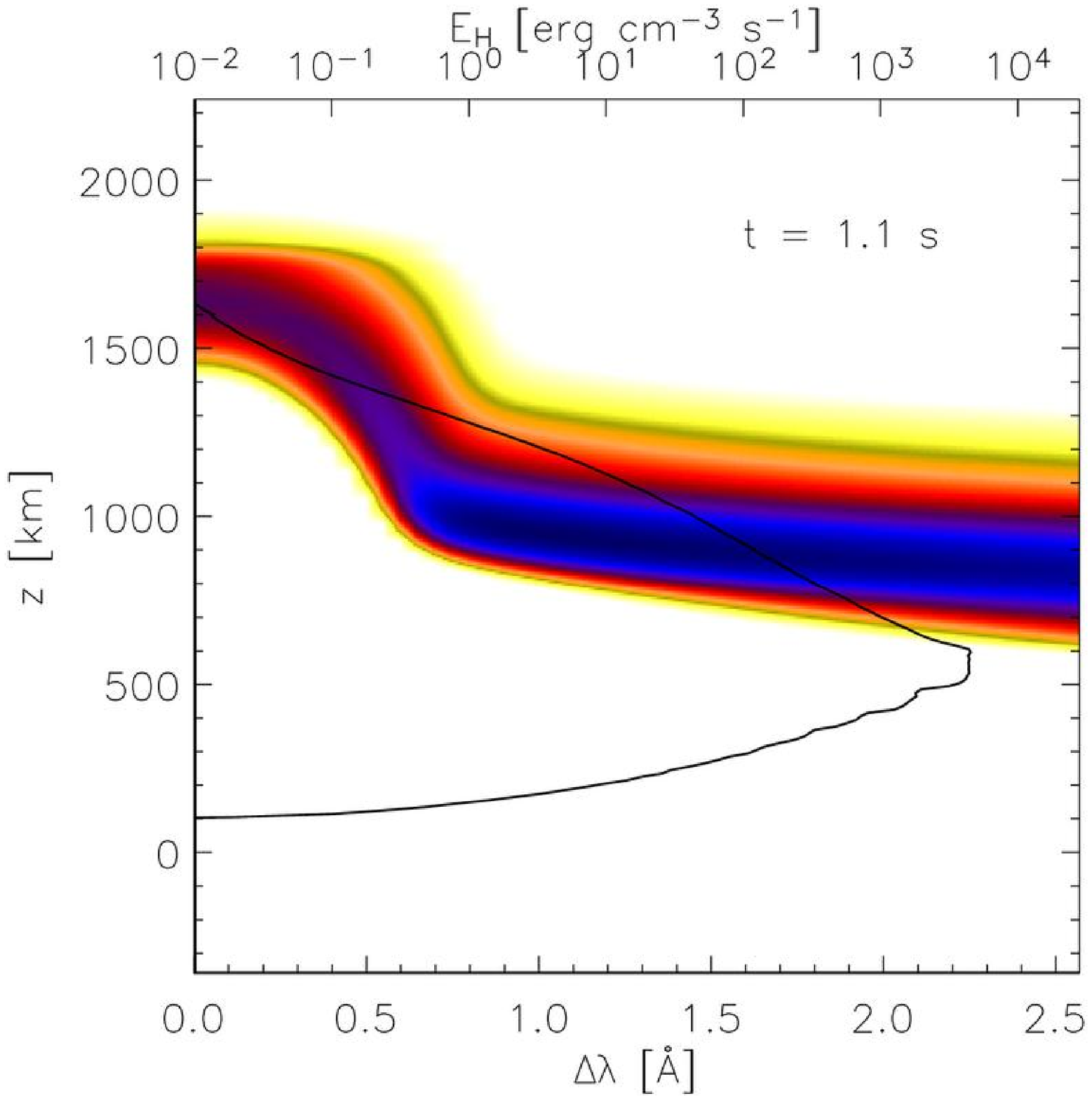}
  \includegraphics[width=0.46\textwidth]{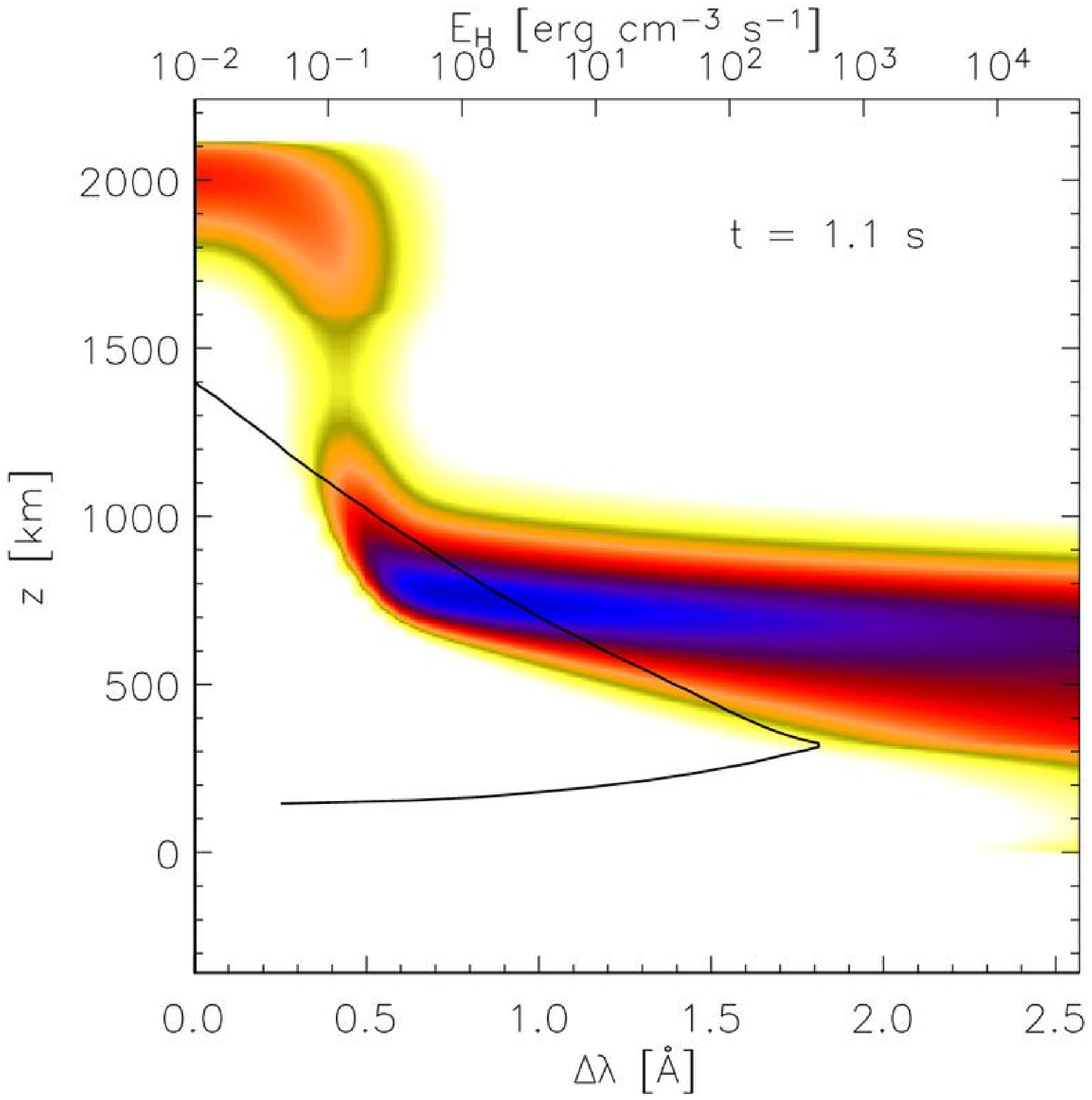}
  \caption[]{\label{kasparova-fig:cfproton} Comparison of H$\alpha$
  contribution function $CF$ for proton beams with $E_1=5$~MeV (first
  column) and $E_1=20$~MeV (second column).  $C^{\rm nt}$ were
  included.  }
\end{figure}

Figure~\ref{kasparova-fig:deposit} shows that RC significantly
increases the energy deposit of electron beams at heights $> 1500$~km
leading to corresponding increase of the temperature and ionization.  The
atmospheric response to proton beams was modelled for two different
values of the low-energy cutoff $E_1=5,\ 20$~MeV.  Note that deka-MeV
protons produce approximately the same amount of hard X-rays as
deka-keV electrons (\cite{kasparova-em85}).
Figure~\ref{kasparova-fig:deposit} illustrates that proton beams
deposit their energy into deeper layers than electron beams
(\cite{kasparova-em96}).  The energy deposit of proton beams with
lower value of $E_1$ peaks at higher atmospheric layers and is larger
at heights above in comparison with a proton beam of $E_1=20$~MeV.
Consequently, the temperature at these layers is substantially
increased.

\section{H$\alpha$ Line Profiles}
The hydrogen level populations are affected also by non-thermal
collisional ionization and excitation.  The corresponding rates
$C^{\rm nt}$ are directly proportional to the beam's energy deposit on
hydrogen (see \cite{kasparova-fhg93} for electron beams, and
\cite{kasparova-hfg93} for proton beams).  Their influence on the
temporal evolution of the H$\alpha$ profile was studied for electron
beams by \citet{kasparova-he91,kasparova-ka05c}.  Here, we describe in
detail their effect for proton beams and the influence of RC for
electron beams.  The temporal evolution of the H$\alpha$ line-center
($\Delta\lambda=0$\,\AA) and wing ($\Delta\lambda=1$\,\AA) intensities
are shown in Fig.~\ref{kasparova-fig:prof}.  As discussed in
\citet{kasparova-he91} and \citet{kasparova-ka05c} for the case of
electron beams, $C^{\rm nt}$ cause a decrease of the line-center
intensity at the very start of the beam propagation and enhance the
wing intensity later on (mainly for fluxes $\gtrsim
10^{11}$~erg\,cm$^{-2}$\,s$^{-1}$).  The line behaviour can be
understood in terms of contribution functions $CF$ to the outgoing
intensity given by $I_\lambda=\int CF_\lambda\,{\rm d}z$, where $z$ is
the height.

Figure~\ref{kasparova-fig:cfel} shows the evolution of H$\alpha$ $CF$
for the case of an electron beam with RC and $C^{\rm nt}$ included.  A
decrease of the line center intensity is caused by an increase of
opacity due to $C^{\rm nt}$ (see $CF$ at 0.2~s).  Later on, a new
region of wing formation occurs at the layers where the energy deposit
peaks (see $CF$ at 0.4~s).  Such a layer is not formed if $C^{\rm nt}$
are not considered.  Since RC causes heating of the top parts of the
atmosphere, it is responsible for the increase of the line center
intensity at 0.4~s (see Fig.~\ref{kasparova-fig:prof}) forming in a
narrow region at $\sim2000$~km (see Fig.~\ref{kasparova-fig:cfel}).

The situation for a proton beam is shown in
Fig.~\ref{kasparova-fig:cfproton}.  In the case of $E_1=20$~MeV, the
beam energy is deposited in regions where the line wings are formed.
If $C^{\rm nt}$ are included, they cause increase of opacity which
leads to a drop of both the wing and the line-center intensities (see
$CF$ at 0.2~s).  Later on, a new wing region occurs.  Neglecting
$C^{\rm nt}$, the line intensity does almost not change because the
temperature structure is not significantly affected by the beam
propagation.

However, for a proton beam of $E_1=5$~MeV, the temperature increase is
large enough to form a new region of the line center intensity, as in
the case of RC for an electron beam.  Also, the energy deposit is
large enough to create a new layer of strong wing intensity (see $CF$
at 0.4 and 1.1~s and Fig.~\ref{kasparova-fig:prof}).  The second peak
of the line center intensity at about 1.1~s is due to broadening of
the formation region.  Similarly to the electron beams, the line wings
are not changed if $C^{\rm nt}$ are not considered.

\section{Conclusions}
The H$\alpha$ line is influenced by temperature structure resulting
from the beam propagation and return current as well as the
non-thermal collisional rates.  We propose to use the decrease of
H$\alpha$ line-center intensity as diagnostic indicating the presence
of particle beams.  We also predict that proton beams with deka-MeV
low-energy cutoffs produce only {\em decrease\/} of the H$\alpha$
in comparison with quiet-sun intensities.\\

\acknowledgements This work was mainly supported by the grants
205/04/0358 and 205/06/P135 (GA CR), partially by grants IAA3003202,
IAA3003203 (GA AS), and by the key project AV0Z10030501 (AI AS).

\end{document}